\documentstyle[aps]{revtex}
\input epsf
\def\beq{\begin{equation}}
\def\eeq{\end{equation}}
\def\gfe2{\gamma -  Fe_2 O_3}

\def\jpc#1#2#3{{\it J. Phys. C: Solid State Phys.} {\bf #1}, #2 #3}

\begin{document}
\title{Composition dependent magnetic properties of iron oxide - polyaniline nanoclusters}

\author{Raksha Sharma, Subhalakshmi Lamba and S. Annapoorni}

\address{ Department of
Physics and Astrophysics, Delhi University, Delhi, India - 110007 
} 
\author {Parmanand Sharma}
\address{Japan Science and Technology Agency, Sendai 980-8577,Japan }
\author{Akihisa Inoue}
\address{Institute for Materials Research, Tohoku University, Sendai 980-8577,Japan }

\maketitle

\begin{abstract}
   $\gfe2$ prepared by sol gel process was used to produce nanocomposites with polyaniline of varying aniline concentrations. TEM shows the presence of chain like structure for lower polyaniline concentration. The room temperature hysteresis curves show finite coercivity of $\sim 160$ Oe for all the composites, while the saturation magnetization was found to decrease with increasing polymer content. ZFC - FC  magnetization measurements indicate high blocking temperatures.
 It is believed that this indicates a strongly interacting system, which is also shown by our TEM results. Monte Carlo simulations performed on a random anisotropy model with dipolar and exchange interactions match well with experimental results.  
\end{abstract}



\section{Introduction}
Nanostructured  materials have attracted a lot of attention due to their commercial exploitation as sensors, batteries, toners in photocopying, quantum electronic devices, smart windows and memory applications. Nanostructured magnetic materials are now being extensively studied for high capacity magnetic storage media, integrated circuits, color imaging, magnetic refrigeration, biomedical application and spintronics \cite{r91,r91n,r92,r93,r93n}.  The agglomeration in these magnetic nanoparticles due to Vander Waal's and magneto static interparticle interaction  hampers their use for technological applications. Several methods have been employed to reduce this problem of agglomeration, namely, blending colloidal nanomagnetic particles with polymers, coating of oxide particles with polymers like PVA, PVP,PMA  etc., forming a core shell structure or embedding the particles in a polymer matrix \cite{r94,r95,r96,r97}.
These so called nanocomposites are also used in blend with conducting polymers like polypyrrole (PPY), polyaniline (PANI) etc.

Iron oxide - conducting polymer  core shell nanocomposites which exhibit  the  properties of both constituents i.e. magnetic as well as conducting, have  proved to be a useful material for many applications like electromagnetic shielding, gas and humidity sensors etc. \cite{r99}.
Magnetic investigations on $\gfe2$ - PPY nanocomposites \cite{r98} reveal that the size and hence the magnetic properties can be controlled by varying the PPY content in the nanocomposite.

The properties of fine particle magnetic systems are known to be strongly dependent on the size, anisotropy and interactions like the exchange and dipolar \cite{r910,r911}. The effect of interparticle interactions in magnetic nanoparticle systems has been investigated by different authors \cite{new2,new3}. In magnetic nanocomposites these factors are easily controlled by varying the composition of the polymer. Hence it is seen that  all  the magnetic properties of these systems  are sensitive to polymer concentration. Our earlier investigations on the $\gfe2$ polypyrrole nanocomposites reveal that superparamagnetic thresholds as evidenced by the presence of blocking are very high even for fairly small particles. Our numerical simulations based on the random anisotropy, interacting model for single domain magnetic particle arrays also show that for a system with exchange interactions, as possible in a nanocomposite, has fairly high blocking even for particles sizes $\sim 10 - 15$ nm \cite{r912}.  

	In this paper we report the preparation of magnetic  nanocomposites of polyaniline and iron oxide.  In order to determine the size, composition, structural and magnetic properties of the nanocomposite several investigations viz. transmission electron microscopy (TEM) Fourier Transform Infrared spectroscopy (FTIR), X-ray diffraction (XRD), hysteresis measurements and the low field (applied field of  $25$ Oe and  $100$ Oe) FC and ZFC magnetization measurements have been carried out.  We also perform Monte  Carlo simulations on the interacting random anisotropy model  using certain  parameters derived from our experimental results , like the particle size and interparticle separation. The purpose of the simulation is to estimate the strength of   typical exchange and anisotropy energies in this system.  

In section \ref{sec2} we describe the method of preparation, in section \ref{sec3} we present the characterization results. In section \ref{sec4} we present the magnetization studies, in section \ref{sec5} we describe the method and results of our simulation studies and in section \ref{sec6} we present our conclusions.

\section{Experimental Synthesis}
\label{sec2}
	The $\gfe2$  nanoparticles were synthesized by sol gel process using ferric nitrate ( $Fe (NO_3)_3 . 9 H_2 O$ )  as precursor and 2-methoxy ethanol as solvent \cite{r98}.
These particles were further dispersed in a mixture of 10 cc deionized water, 50 cc of 1M $HCl$ and .4 cc (.39 gm) of aniline monomer. The above solution was continuously stirred for 1 hour at ice temperature before adding the oxidant. Agitation due to stirring reduces the agglomeration by breaking the big clusters into smaller ones. On addition of oxidant prepared by dissolving ammonium persulphate (1 gm) in 20 ml distilled water at a rate of .08 ml/min,  simultaneous polymerization and stirring leads to coating of PANI chains on these smaller $\gfe2$ particles.
 To obtain coating of polyaniline  1 gm of $\gfe2$ was dispersed  in a mixture of $10$ cc deionized water and $10$ cc of $1$ M HCl. To the above dispersion  $0.4$ cc ($0.39$ gm) of aniline monomers were added.  This solution was maintained at $ 0-50$ C and constantly stirred for 1 hour. A  solution containing $1$ gm of ammonium persulphate in $20$ cc of deionized water was added  to the above.  The reaction was continued for 4 hours with the solution placed in an ice water bath, after which it  was  filtered and washed repeatedly  with $1$ M  HCl till the disappearance of the color of the filtrate. Subsequently it was washed with methanol and diethyl ether.  The powder  was allowed to dry  at room temperature  for 48 hours.  It was observed that the color of the powder thus obtained for different concentrations of $\gfe2$ : PANI (like $1:0$, $1:0.05$ ,$1:0.1$ , $1:0.2$ and $1:0.4$ )  changed from  brown to dark green   as  the concentration of aniline monomer was increased. This shows the presence of a higher concentration of the polymer in the composite since pure polyaniline in its doped form is dark green in color.

	The crystal structure of the powder was examined by a Rigaku Rotaflex diffractometer using a $Cu -  k \alpha$ ($\lambda = 1.5918$ A)   at 40 KeV.  The size and shape of the particles were analyzed by a JEOL JEM 2000 EX TEM. The presence of the polymer was confirmed  using a Nicolet 510 P FTIR spectrometer. A Quantum Design MPMS-5S Quantum Design SQUID magnetometer was used for magnetic characterisation in the temperature range of $5$ K to $310$ K.  For the zero field cooled (ZFC) measurements the sample was cooled down to $5$ K in the absence of an external magnetic field followed by measurements in a constant magnetic field during the warming run of the experiments.  In the field- cooled(FC) measurments, the sample was cooled from $310$ K down to $5$ K in the presence of the same constant magnetic field which was also used in the subsequent warm up scan.  The room temperature ($\sim 300 $ K) hysteresis loop  M(H) was measured using a vibrating sample magnetometer (VSM-5, TOEI Industry Co. Ltd. Tokyo, Japan), and the samples were subjected to a magnetic field cycling between $+1.5$ Tesla and $-1.5$ Tesla.  
\section{Results and Discussion}
\label{sec3}
\subsection{X-Ray Diffraction}
	Fig.1 (a) - (d) show XRD pattern of different compositions of the $\gfe2$  : PAN1  composite, unannealed, in varying ratios viz. 1:0, 1:0.1, 1:0.4 and 0:1 (Pure PANI). Peaks were observed at $2 \theta = 30.360, 33.982, 35.691$  and $57.400$, with hkl values ($206$), ($109$), ($119$) and  ($11{\overline 15}$) which corresponds to the $\gfe2$ phase.  The lattice constant was calculated to be a = 8.3452 $\pm$ .0001 Awhich compares well with the literature value of  8.34 A \cite{new1}.  It was observed that with increase in concentration of polyaniline the intensity of peaks decreases and  also some low intensity peaks are suppressed. This can be explained by the thicker coating of PANI on iron oxide which  suppresses the  peaks at higher concentrations.  But it can be seen that the  same phase of iron oxide is retained in all the concentrations. For pure PANI an amorphous nature is observed.

\subsection{Transmission Electron Microscopy}
	The transmission electron micrograph TEM in Fig.2(a) of unannealed iron oxide shows  agglomerated clusters of iron oxide. In the presence of polymer  the particle size reduces drastically as can be seen from Fig.2(b) which is a composite in the ratio $1:0.1$ i.e. a low concentration of polymer.
The presence of polymer prevents agglomeration of the iron oxide and hence smaller clusters are observed for composites.
 There is still some agglomeration  present  due to the partial coating of the  polymer. As the polymer concentration is doubled  ($1:0.2$) the $\gfe2$ particles were found to align in a chain as seen in Fig. 2(c).  On further increasing polymer concentration intrachain structure starts building up, which is seen in Fig. 2(d) which has a composition of $1:0.4$  . However the size of the particle is found to reduce further.  

\section{Magnetization measurements}
\label{sec4}
In Figure 3 we present the hysteresis measurements performed at room temperature for three different compositions. It is seen that the magnetic nanocomposites show a finite coercivity at  room temperature for all the concentrations. This indicates that the system is still in the ferromagnetic regime, in spite of the small particle sizes. All the three systems show a tendency to saturate at $1.5$ Tesla and the coercivities are in the range of $160$ Oe.  The saturation magnetization $M_S$  are found to be  53:37; 44.29; and 35.31 emu/gm  for concentrations of  1:0.1, 1:0.2 and 1:0.4  respectively, while the corresponding retentivities are  $\sim 8.5$, $7$ and $5.9$   $emu/gm$.
The reduced value of the saturation magnetization for increasing polymer concentration is expected, due the presence of smaller quantities of $\gfe2$ in the sample. In the inset we have plotted the same data in the field range of $-2$ to $2$ K. Oe. The value of the coercivity is almost the same for all the concentrations indicating that the anisotropy energy of the magnetic particles is not much affected by the presence of polymer. The low value of the coercivity indicates that the system is close to superparamagnetic behavior.

In order to confirm the onset of the superparamagnetic phase, ZFC -FC magnetization measurements were performed for one of the above samples viz 1:0.4 for which the particle sizes  seem to be very small (Fig. 2(d)) and  the results are shown in Figure 4 .  The measurements are performed at two different field strengths of 25 (curve 
(a)) and 100 Oe (curve (b)). At 100 Oe the system seems to be closer to blocking at 300 K than at 25 Oe.  
It is our belief that for systems of  magnetic oxide nanoparticles which have a tendency to cluster, the blocking temperatures is not only decided by the particle size and anisotropy constant but also the exchange and dipolar interactions. Large exchange interactions have a tendency to increase the blocking temperature, which we feel may be the case in these systems.

\section{ Simulation and Results}
\label{sec5}
The simulation is performed with an array of $N$ single domain magnetic particles positioned randomly in a cube of side $L$. Each particle has a magnetic moment vector  $\vec \mu_i$ and the direction of the easy axis of magnetization of the particle is represented by the unit vector $\vec n_i$.The  magnetic moment vector  for a single particle has a temperature independent  value  $ \vec \mu_i = V_i M_S \vec \sigma_i$ where  $M_S$ is the saturation magnetization and $\vec \sigma_i $ is the unit vector along the direction of magnetization. The volume of the particle  $V_i$ is picked from a normal distribution $P(V)dV ={1\over (2\pi t ^2)^1/2 }\exp \left(- {(V-V_0) ^2 \over 2 t^2}\right) $ where $V_0$ represents the mean volume  of the particle and $t$ is the width of the distribution which  is taken to be $.1$. This is to account for the experimental results which show that the  magnetic nanoparticles in the sample are (i) not all of the same shape and  size and (ii) positioned randomly in the sample.
The directions of the easy axis of magnetization of each particle is  picked randomly, keeping in mind that the particles could have different shapes. Such an array  of interacting single domain magnetic nanoparticles can be described using the following  Hamiltonian, \cite{r19},
\begin{eqnarray}
H&=&- K \sum_i V_i { \left (\vec \mu_i . \vec n_i  \right )^2 \over |\vec \mu_i|^2} - \sum _{<i\neq j>}J _{ij}\vec \mu_i . \vec \mu_j\nonumber \\
&-& \mu _0  \sum_{<i\neq j>}
{3 (\vec \mu_i.\vec e_{ij})( \vec \mu_j.\vec e_{ij})- \vec \mu_i. \vec \mu _j \over r_{ij}^3} \nonumber \\
 &+& \mu_0  H  \sum _j  \vec \mu_j\nonumber \\
\end{eqnarray}
\noindent The first term in Eq. 1 represents the anisotropy energy of the $i$ th  magnetic particle which has  an  anisotropy constant  $K$. The second term is the exchange interaction  energy between the  different particles in the array and  $J_{ij}$ is the strength of the  ferromagnetic exchange interaction between two  particles with localized magnetic moment vectors   $\vec \mu_i$ and $\vec \mu_j$ respectively. The third term  is the dipolar interaction between these particles, with $r_{ij} $ as the distance between the i$th$ and j$th$ particles and  $\vec e_{ij} $  the unit vector pointing along $\vec{r_{ij}}$. The last term is the energy of the particles due to an externally applied magnetic field $H$.  For the purpose of simulation we  assume that the exchange interaction has a site independent constant value $J_{eff}= J $.  

Magnetization studies ( FC-ZFC)  show that the blocking temperatures are very high, which is evident even in the hysteresis measurements. Our earlier simulations \cite{r912} suggests that the shape of the hysteresis curve (Fig.3 ) is typical of a system in which both exchange and dipolar interactions play a role. It is not possible to derive the value of either $K$ or $J$ from any of our experimental results. So for our simulation we use as an input the particle size and the typical interparticle distance,  and try to obtain the experimentally observed  coercivity and retentivity by varying the value of $K$ and $J$.  
We work with a system of   $N= 64$  particles. The mean volume $V_0$ is taken to be equivalent to that of a sphere  of radius $r_i=7.5 $ nm.  The dipolar interaction energy is calculated by summing over periodic repeats of the basic simulation cell by the method of Lekner summation \cite{r21}. The simulation of the hysteresis is performed at a temperature of $300$ K  by the Monte Carlo method using the standard Metropolis algorithm \cite{r912,r10,r23}.
The simulation is started at a very low field  along the $z$- axis of the simulation cube, using an ensemble of magnetization vectors  for the array, which gives  a  net magnetization of  zero along the $z$ - direction.
 The system is saturated by gradually increasing the magnetic field up to a very high field which is sufficiently higher than its anisotropy field (in this case 1.5 Tesla). Then the magnetic hysteresis loop is simulated.  For each value of the magnetic field $10000$ Monte Carlo  steps are used for the thermalization of the system and the calculation of the net magnetization along the field direction is made over the next $5000$ Monte Carlo  steps. The results are averaged over five initial configurations of the magnetic particles(position, magnetization and easy axes directions). 
 The values of $M_S =4 \times 10^ 5 $ Amp/m  for $\gamma -  Fe_2 O_3$ is  taken from literature. To fit to the experimentally observed values of the coercivity and retentivity  in these systems,  we find that the anisotropy constant should be higher than for the pure $\gfe2$ system which is $\sim .045  \times 10 ^5$ $J/m^3$. A good fit to the experimental results are obtained for a much higher value of the anisotropy constant which is  $K = .75\times 10^5 $  $J/m^3$  and a  value of  $J =.02 E_A$, where  the mean anisotropy energy  $E_A = K V_0$. The values of the anisotropy constant calculated from ac susceptibility measurements performed on coated superparamagnetic iron oxide  nanoparticles shows that the values of $K$ are sensitive to the coating material, and for small particles of $\gfe2$ and $Fe_3 
0_4$  can be an order of magnitude larger than the values for bulk, in fact   high as  $2 - 4 \times 10 ^5 $ $J/m^3$ \cite{a1,a2}. So the value of $K$ estimated from the simulation is well within the experimentally reported limits for coated experimental oxides.
In Figure 5. we show the results of the hysteresis simulations, by plotting the scaled magnetization $M/M_S$ with the applied field, for three different values of the anisotropy constant keeping $V_0$, the interparticle distance and the exchange interaction same for all three curves. Curve (a) is the experimental plot at $300$ K for the $\gfe2$ - PANI concentration of $1:.4$. It is seen that the closest fit to the experimental curve is for a value of  $K=.75 \times 10^5$ $J/ m^3$ which is curve(c). For $K=1.25 \times 10^5$ $J/m^3$  which is curve(d), the coercivity is much larger than that observed experimentally  and for  $K=.25 \times 10^5$ $J/m^3$ (curve (d)) the coercivity is $0$ indicating the system is already superparamagnetic at $300$ K.  

Using the same  parameters and the value of $K=.75 \times 10^5$ $J/ m^3$ which is closest to the experimental curve,  we simulate  the FC- ZFC magnetization curves which are shown in Fig. 6. The effect of  a large $K$ is that the blocking temperatures $\sim 325 $ K  for $H=100$ Oe, which again is in good agreement with the experimental result (Fig. 4), indicating that the values of $K$ and $J$ estimated from the hysteresis  simulation are appropriate for the system being studied. The reasons for the high blocking temperatures even for such small particles are  the presence of  interactions. The broadness of the ZFC peak is indicative of the size distribution in the sample.

\section{Conclusions}
\label{sec6}
Our experiments indicate that  for fairly low concentrations of polyaniline in the nanocomposite, we are able to obtain large variation in the particle size and magnetic properties. The particle sizes are small and  even at such low concentrations of polyaniline there is less agglomeration. High blocking temperatures as confirmed by hysteresis and ZFC - FC  magnetization measurements indicates that the system is in the ferromagnetic phase at room temperature. However the low value of coercivity suggests that the system is close to superparamagnetic behaviour. We are further investigating the electrical properties, like the conductivity and the dielectric behaviour  of these nanocomposites to understand the contribution of the polymer in the system.

\noindent {\bf{Acknowledgements}}
One of us (RS) wishes to acknowledge CSIR for financial support. We also acknowledge the help extended by Dr. N. C. Mehra and Dr. S.K. Shukla at the University Science and Instrumentation Centre, University of Delhi.

\begin{figure}
\caption{X- Ray Diffraction for (a) pure unannealed $\gfe2$  and two  different compositions of $\gfe2$ - PANI nanocomposites which are (b)$ 1:.1$ and  (c) $1:.4$, and (d) pure PANI}
\end{figure}

\begin{figure}
\caption{TEM  for (a) pure unannealed $\gfe2$  and three different compositions of $\gfe2$ - PANI nanocomposites which are (b)$ 1:.1$  (c) $1:.2$ and (d) $1:.4$.}
\end{figure}

\begin{figure}[ht]
\epsfbox{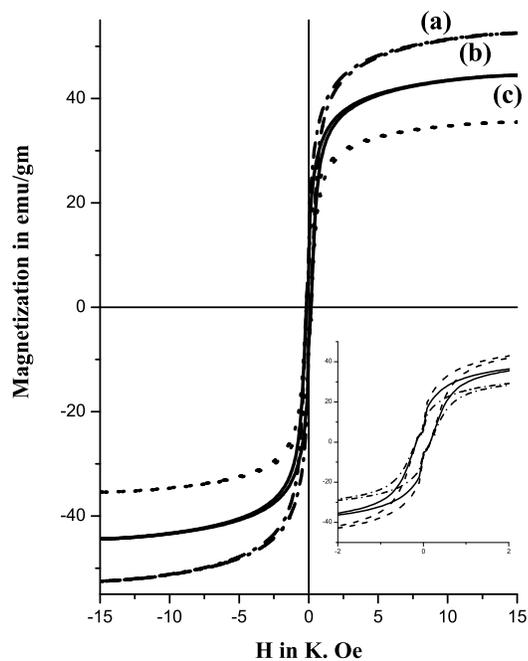}
\caption{ Experimental hysteresis at $300$ K for three different compositions of $\gfe2$ - PANI nanocomposites which are  (a)$ 1:.1$  (b) $1:.2$ and (c) $1:.4$. Inset shows the data plotted over the low field range}
\end{figure}

\begin{figure}[ht]
\epsfbox{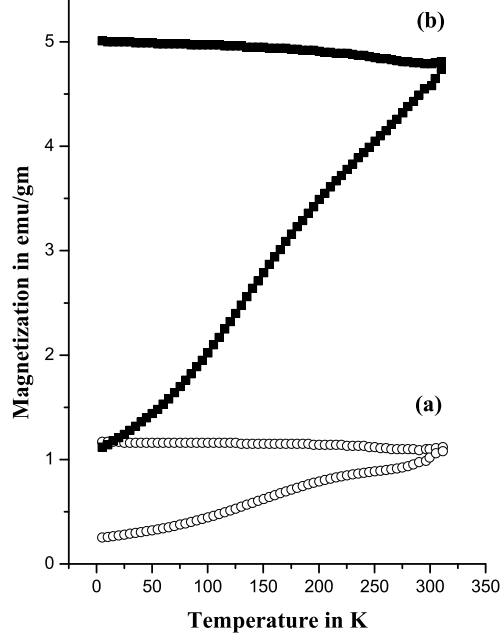}
\caption{Experimental  ZFC - FC magnetization results for $1:.4$  composition of  $\gfe2$ - PANI nanocomposite at  (a)$25$ Oe and  (b) $100$ Oe..}
\end{figure}

\begin{figure}[ht]
\epsfbox{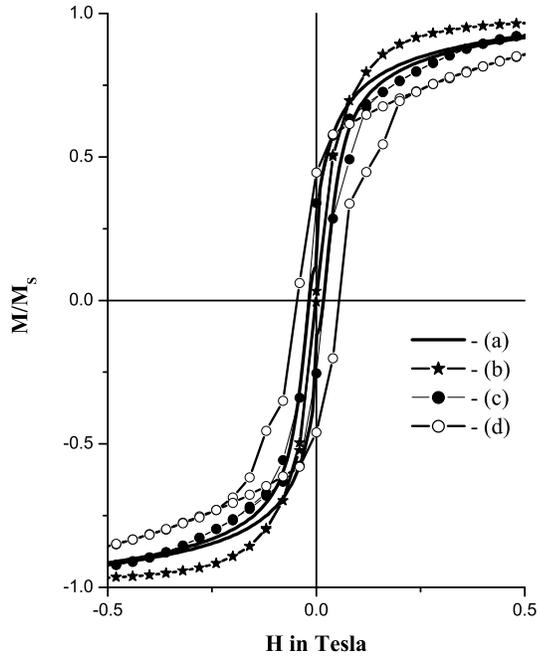}
\caption{(a) Experimental hysteresis at $300$ K for  1:0.4 composition of $\gfe2$ - PANI composition and simulated hysteresis curves at 300 K for $J=.02 E_A$ and  (b)$ K = .25 \times 10^5$  $J/m^3$ , (c)$ K = .75 \times 10^5$  $J/m^3$ and (d) $ K = 1.25 \times 10^5$  $J/m^3$.}
\end{figure}
\begin{figure}[ht]
\epsfbox{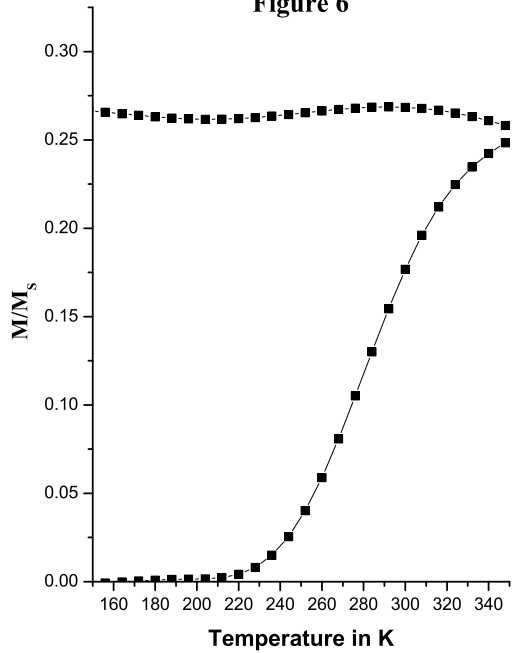}
\caption{(a)Simulated ZFC - FC magnetization  curves for $J=.02 E_A$ , $ K = .75 \times 10^5$  $J/m^3$ and  $H=.01$ Tesla}
\end{figure}


\begin{thebibliography}{50}

\bibitem{r91}
F. J. Himpsel,\jpc{11} {9483}  {(1999)}
\bibitem{r91n}
P. Moriarty, {\it Rep. Prog. Phys.} {\bf 64} {297} {(2001)}
\bibitem{r92}
N. M. White and J. D. Turner, {\it Meas. Sci. Technol.} {\bf 8} 1 (1997)
\bibitem{r93}
D. A. Thompson and J. S. Best, {\it IBM J. Res. Develop.} {\bf 44} 311 (2000)
\bibitem{r93n}
P. Sharma, A. Gupta, K. V. Rao, F. J. Owens, R. Ahuja, J. M. O. Guillen, B. Johansson, and G. A. Gehring,{\it  Nature Materials }, {\bf 2} 673 (2003).
\bibitem{r94}
Lei Chen, Weng-Jun Yang and  Chang-Zheng Yang,{\it. J. Mater. Sci.} {\bf 32} 3571 (1997)
\bibitem{r95}
D. K. Lee, Y. S. Kang, C. S. Lee and P. Stroeve, {\it J. Phys. Chem. B} {\bf 106} 7267 (2002)
\bibitem{r96}
 S. Maeda and S. P. Armes,{\it J. Mater. Chem} {\bf 4} 935 (1994)
\bibitem{r97}
C. L. Huang and E. Matijevic,{\it J. Mater. Res.} {\bf 10} 1327 (1995)

\bibitem{r99}
K. Suri, S. Annapoorni,A. K. Sarkar and   R. P. Tandon, {\it Sensors and Actuators B} {\bf 81} 277 (2002)
\bibitem{r98}
K. Suri, S. Annapoorni, R. P. Tandon and N. C. Mehra,{\it  Synth. Metals} {\bf 126} 137 (2002) 

\bibitem{r910}
Magnetic Properties of Fine Particles, Ed. J. H. Dormann and D. Fiorani, (North Holland, Amster dam), (1992)
\bibitem{r911}
 D. Fiorani,  J. H. Dormann, R. Cherkooui, E. Tronc, F. Lucari, F. D. Orazio, L. Spinu, M. Nogues , A. Garcia and A. M. Testa, {\it J. Mag. Mag. Mater} {\bf 196 - 197}143 (1999) 
\bibitem{new2}
C. Papusoi Jr.,{\it J. Mag. Mag. Mater} {\bf 195}708 (1999)  
\bibitem{new3}
D. Hinzke and U. Nowak, Phys. Rev. B  {\bf 61},  6734 (2000)
\bibitem{r912}
S. Lamba and S. Annapoorni,  {\it Euro. Phys. Jour. B} {\bf 39} 19 (2004) 
\bibitem{new1}
International Centre for Diffraction Data (1979),Inorganic Vol. PDIS - 20iRB, Publ. JCPDS
\bibitem{r19}
J.J Weis, {\it J. Phys. Cond. Matter} {\bf 15} S1471 (2003)
\bibitem{r21}
John Lekner,{\it J. Phys. A} {\bf 176} 485 (1991)
\bibitem{r10}
L. Wang, J. Ding, H. Z. Kong, Y. Li and Y. P.Feng, {\it Phys. Rev. B} {\bf 64} 214410 -1 (2001)
\bibitem{r23}
K. Binder, H. Rauch and V. Wildpaner, {\it J. Phys. Chem. Solids} {\bf 31} 391 (1970). 
\bibitem{a1}
S. Morup and H. Topsoe, {\it Appl. Phys.} {\bf 11}  63 (1976)
\bibitem{a2}B. J. Jonsson, T. Turkki, V. Strom, M. S. El - Shall and K. V. Rao, {\it J. Appl. Phys.} {\bf 79} 5063 (1996)
\end{thebibliography}
\end{document}